\documentclass{article}

\usepackage{arxiv}

\usepackage[utf8]{inputenc} % allow utf-8 input
\usepackage[T1]{fontenc}    % use 8-bit T1 fonts
\usepackage{hyperref}       % hyperlinks
\usepackage{url}            % simple URL typesetting
\usepackage{booktabs}       % professional-quality tables
\usepackage{amsfonts}       % blackboard math symbols
\usepackage{nicefrac}       % compact symbols for 1/2, etc.
\usepackage{microtype}      % microtypography
\usepackage{lipsum}		% Can be removed after putting your text content
\usepackage{graphicx}
\usepackage{natbib}
\usepackage{doi}

\usepackage{tikz}

\usepackage{amsmath}
\usepackage{bm}
\usepackage{caption}
\usepackage{subcaption}
\usepackage[cmintegrals]{newtxmath}

\begin{document}
\title{Scheduling Algorithms for Age of Information Differentiation with Random Arrivals}
	\author{
		Nail~Akar\\
		Electrical and Electronics Engineering Dept.\\
		Bilkent University, 06800\\
		Ankara, Turkey \\
		\texttt{akar@ee.bilkent.edu.tr} \\
		\And
		Ezhan~Karasan\\
		Electrical and Electronics Engineering Dept.\\
		Bilkent University, 06800\\
		Ankara, Turkey \\
		\texttt{ezhan@ee.bilkent.edu.tr} \\
	}
	\maketitle

\maketitle

\begin{abstract}
We study age-agnostic scheduling in a non-preemptive status update system with two sources sending time-stamped information packets at random instances to a common monitor through a single server. The server is equipped with a waiting room holding the freshest packet from each source called "single-buffer per-source queueing". The server is assumed to be work-conserving and when the waiting room has two waiting packets (one from each source), a probabilistic scheduling policy is applied so as to provide Age of Information (AoI) differentiation for the two sources of interest.
Assuming Poisson packet arrivals and exponentially distributed service times, 
the exact distributions of AoI and also Peak AoI (PAoI) for each source are first obtained.
Subsequently, this analytical tool is used to numerically obtain the optimum probabilistic scheduling policy so as to minimize the weighted average AoI/PAoI by means of which differentiation can be achieved between the two sources. In addition, a pair of heuristic age-agnostic schedulers are proposed on the basis of heavy-traffic analysis and comparatively evaluated in a wide variety of scenarios, and guidelines are provided for scheduling and AoI differentiation in status update systems with two sources.
\end{abstract}

% \fi

\section{Introduction}
% \IEEEPARstart{A}{demo} file is ....\cite
\label{section1}
Timely status updates play a key role in networked control and monitoring systems. 
Age of Information (AoI) has recently been introduced to quantify the timeliness of information freshness in status update systems \cite{kaul_etal_SMAN11}.  
In the general AoI framework outlined in \cite{kosta_etal_survey}, information sources sample a source-specific random process at random epochs and generate information packets containing the sample values as well as the sampling times.
On the other hand, servers gather the information packets from multiple sources so as to be transmitted to a remote monitor using queueing, buffer management, scheduling, etc. 
For a given source, AoI is defined as the time elapsed since the generation of the last successfully received update packet. Therefore, AoI is a source-specific random process whose sample paths increase in time with unit slope but are subject to abrupt downward jumps at information packet reception instances. The PAoI process is obtained by sampling the AoI process just before the cycle termination instances.

\begin{figure}[tb]
	\centering
	\begin{tikzpicture}[scale=0.32]
	%\draw[ultra thick, blue] (182) circle (2) ;
	\draw[very thick](3,2) circle (2);
	\draw (3,4) node[anchor=south] {\small{source-$2$}} ;
	%\draw[ultra thick, blue] (2,12) circle (2)  ;
	\draw[very thick] (3,9) circle (2)  ;
	\draw (3,11) node[anchor=south] {\small{source-$1$}} ;
	\draw[very thick,->] (6,2) -- (10,2) ;
	\draw (8,2) node[anchor=south] {$\lambda_2$};	
	\draw[very thick,->] (6,9) -- (10,9) ;
	\draw (8,9) node[anchor=south] {$\lambda_1$};	
	\filldraw[fill=gray!50, thick] (13,1) rectangle(15,3);
	\draw[thick] (11,1) -- (13,1);
	\draw[thick] (11,3) -- (13,3);
	\filldraw[fill=gray!50, thick] (13,8) rectangle(15,10);
	\draw[thick] (11,8) -- (13,8);
	\draw[thick] (11,10) -- (13,10);
	\draw[thick,->] (15,2) -- (20,5) ;
	\draw[thick,->] (15,9) -- (20,6) ;
	\draw[thick, ->, dashed] (20,3.5) arc (270:90:2) node[anchor=south east] {$p_1$};
	\filldraw (20,3.5) circle (0.01) node[anchor=north east] {$p_2$};
	\draw[very thick](24,6) circle (3);
	\filldraw (24,6) circle (0.01) node[anchor=center] {server};
	\draw[very thick,->] (28,6) -- (32,6) ;
	\draw[very thick](33,4) rectangle (40,8);
	\filldraw (36.5,6) circle (0.01) node[anchor=center] {monitor};
	%\draw[ultra thick, blue] (2,18) circle (2)  ;	
	%	\fill[blue] (2,18) circle (2);
	%	\draw (2,20) node[anchor=south] {\small{source-$1$}} ;	
	%	\fill[blue] (15,10) circle (4) node[text=white] {\begin{tabular}{c} 
	%		\small{\text{bufferless}} \\ \small{\text{server}} \end{tabular}} ;
	%	\draw[ultra thick,->] (5,17.5) -- (9,15) ;
	%	\draw (7,16.5) node[anchor=south] {$\lambda_1$};		
	%	\draw[ultra thick,->] (5,12) -- (9,11) ;
	%	\draw (7,11.5) node[anchor=south] {$\lambda_2$};
	%	\draw[ultra thick,->] (5,2.5) -- (9,5) ;
	%	\draw (7,4) node[anchor=south] {$\lambda_N$};
	%	\filldraw[blue] (2,6.5) circle (3pt);
	%	\filldraw[blue] (2,7.5) circle (3pt) ;
	%	\filldraw[blue] (2,8.5) circle (3pt); 
	%	
	%	\filldraw[black] (6.5,7) circle (3pt);
	%	\filldraw[black] (6.5,8) circle (3pt) ;
	%	\filldraw[black] (6.5,9) circle (3pt); 
	%	\draw[ultra thick,->] (5,2.5) -- (9,5) ;
	%	\draw[ultra thick,->] (21,10) -- (25,10) ;
	%	%\draw[ultra thick, blue] (27,8) rectangle (33,12) node {monitor};
	%	\fill[blue] (27,7) rectangle (35,13); 
	%	\draw (31,10) node[text=white] {\begin{tabular}{c} 
	%		\small{	\text{remote}} \\ \small{\text{monitor}} \end{tabular}} ;
	\end{tikzpicture}
	\caption{A single-hop status update system with two sources employing per-source queueing. A single-buffer queue is dedicated to each source to hold the freshest packet from that source.}
	\label{fig:twosource}
\end{figure}
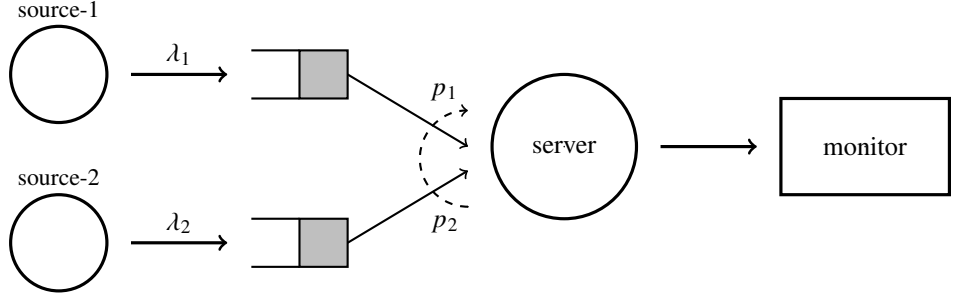

In this paper, we consider a non-preemptive status update system in Fig.~\ref{fig:twosource} with two sources, a server, and a monitor, with random information packet arrivals from the sources.
The server employs Single-Buffer Per-Source queueing (SBPSQ) for which the freshest packet from each source is held in a single buffer. The server is work-conserving, i.e., it does not idle unless the waiting room is empty, and it  serves the packets probabilistically (probabilities denoted by $p_i$ in Fig.~\ref{fig:twosource}) when there are two waiting packets. The scheduler is age-agnostic and does not require the server to read the timestamp field in the packets and keep track of the instantaneous AoI values. The scheduling probabilities are to be chosen so as to provide AoI/PAoI differentiation. In this paper, we attempt to provide differentiation through the minimization of the weighted average AoI/PAoI. The motivation behind AoI differentiation is that in  a networked control system, the information about certain input processes need to be kept relatively fresher at the control unit since this information will have profound impact on the overall performance of the control system.
The studied model falls in the general framework of status update systems analytically studied in the literature; see the surveys on AoI  \cite{kosta_etal_survey},\cite{survey_Yates} and the references therein for a collection of multi-source queueing models for AoI.

Our main contributions are as follows:
\begin{itemize}
	\item As the main contribution of this paper, under the assumption of Poisson packet arrivals, and exponentially distributed service times,  we obtain the exact distributions of the AoI/PAoI processes for the two sources for the system of interest in Fig.~\ref{fig:twosource} as a function of the scheduling probabilities. The analysis is based on well-known absorbing Continuous-Time Markov Chains (CTMC) and does not employ relatively more specific tools such as Stochastic Hybrid Systems (SHS) that were recently proposed for AoI modeling \cite{yates_kaul_tit19} and used in several works. We believe that the simplicity of the tool we use to derive the AoI/PAoI distributions makes it a convenient tool for researchers and practitioners in this field. Subsequently, for given traffic parameters, the proposed analytical model is used to numerically obtain the Optimum Probabilistic Scheduling (OPS) policy which minimizes the weighted average AoI or PAoI, referred to as OPS-A and OPS-P, respectively.
	\item A heavy-traffic analysis is presented to obtain closed-form expressions for the average per-source AoI/PAoI values which has enabled us to write the OPS-P policy in closed-form in heavy-traffic regime. On the other hand, the OPS-A policy for the heavy-traffic regime is shown to be obtainable by solving a quartic equation. 
	 \item On the basis of the heavy-traffic analysis, we propose two age-agnostic heuristic schedulers that are quite easy to implement in comparison with age-aware schedulers and therefore they can be used in more challenging multi-hop scenarios and resource-constrained servers.
\end{itemize}

The paper is organized as follows. Section~\ref{section2} presents the related work. In Section~\ref{section3}, the analytical model is presented. The heavy-traffic regime is addressed in Section~\ref{section4} along with the two heavy-traffic analysis-based heuristic schedulers. Section~\ref{section5} addresses the analytical model and associated closed-form expressions for the Non-Preemptive Bufferless (NPB) variation of the same problem which is used as a benchmark in the numerical examples. In Section~\ref{section6}, we provide numerical examples for comparative evaluation of the age-agnostic schedulers of interest. We conclude in Section~\ref{section7}. 
\section{Related Work}
\label{section2}
There has been a great deal of interest on AoI modeling and optimization problems in the context of communication systems since the reference \cite{kaul_etal_infocom12} first introduced the AoI concept in a single-source, single-server queueing system setting. The existing analytical models can be classified according to one or more of the following: (i) existence of one, two, or more  information sources, (ii) random access vs. scheduled access, (iii) existence of transmission errors, (iv) performance metrics used, e.g., average AoI/PAoI values, age violation probabilities, etc., (v) buffer management mechanisms, (vi) scheduling algorithms, (vii) arrival and service processes used in the models, (viii) single-hop vs. multi-hop systems, (ix) continuous-time vs. discrete-time systems. The recent references \cite{kosta_etal_survey} and \cite{survey_Yates} present exhaustive surveys on existing work on AoI and moreover describe several open problems.

\subsection{Single-source Queueing Models} The average AoI is obtained for the M/M/1, M/D/1, and D/M/1 queues with infinite buffer capacity and FCFS (First Come First Serve) in \cite{kaul_etal_infocom12}. 
The reference \cite{costa_etal_TIT16} obtains the AoI and PAoI distributions for small buffer systems, namely M/M/1/1 and M/M/1/2 queues, as well as the non-preemptive LCFS (Last Come First Serve) M/M/1/2$^{\ast}$ queue for which the packet waiting in the queue is replaced by a fresher packet arrival.
The average AoI and PAoI are obtained in \cite{najm_nasser_isit16} for the preemptive LCFS M/G/1/1 queueing system 
where a new arrival preempts the packet in service and the service time distribution is assumed to follow a more general gamma distribution.
Average PAoI expressions are derived for an M/M/1 queueing system with packet transmission errors with various buffer management schemes in \cite{chen_huang_isit16}.
Expressions for the steady-state distributions of AoI and PAoI are derived in \cite{inoue_etal_tit19} for a wide range of single-source systems.
The authors of \cite{akar_etal_tcom20} obtain the exact distributions of AoI and PAoI in bufferless systems with probabilistic preemption and single-buffer systems with probabilistic replacement also allowing general phase type distributions to represent interrarival times and/or service times. 
%A discrete-time queueing model with Bernouilli arrivals and geometric service times, using FCFS and non-preemptive LCFS scheduling is presented in \cite{kosta_etal_isit19} with expressions for the mean AoI and PAoI values.
%In addition to exact methods, a number of studies provide bounds for certain AoI-related metrics of interest.
%The reference \cite{soysal_ulukus_unpublished} derives upper bounds for the mean AoI for
%the $G/G/1/1$ queue as well as its preemptive version while showing that the bounds are close to actual values. Similarly, the authors of \cite{champati_etal_infocom19}  present a method for obtaining upper bounds for the AoI violation probability for both $GI/GI/1/1$ and $GI/GI/1/2^{\ast}$ systems, in addition to some exact closed-form expressions for some sub-cases. 
\subsection{Multi-source Queueing Models}
For analytical models involving multiple sources, the average PAoI for M/G/1 FCFS and bufferless  M/G/1/1 systems with heterogeneous service time requirements are derived in \cite{huang_modiano} by which one can optimize the information packet generation rates from the sources. 
An exact expression for the average AoI for the case of multi-source M/M/1 queueing model under FCFS scheduling is provided in \cite{moltafet2020average} and three approximate expressions are proposed for the average AoI for the more general multi-source M/G/1 queueing model. 
The reference \cite{yates_kaul_tit19} investigates the multi-source M/M/1 model with FCFS, preemptive bufferless, and non-preemptive single buffer with replacement, using the theory of Stochastic Hybrid Systems (SHS) and obtain exact expressions for the average AoI. Hyperexponential (H$_2$) service time distribution for each source is considered in \cite{yates_etal_isit19}  for an M/H$_2$/1/1 non-preemptive bufferless queue to derive an expression for the average per-source AoI per class.  
The authors of \cite{farazi_etal_Asilomar19} study a self-preemptive system in which preemption of a source in service is allowed by a newly arriving packet from the same source and AoI expressions are derived using the SHS technique.  
For distributional results, the MGF (Moment Generating Function) of AoI has been derived for a bufferless multi-source status update system using global preemption \cite{moltafet2021moment}. The work in \cite{abdelmagid2021closedform} considers a real-time status update system with an energy harvesting transmitter and derive the MGF of AoI in closed-form under certain queueing disciplines making use of SHS techniques.
The authors of \cite{dogan_akar_tcom21} obtain the exact distributions of AoI/PAoI in a probabilistically preemptive bufferless multi-source M/PH/1/1 queue where non-preemptive, globally preemptive, and self-preemptive systems are investigated using a common unifying framework. In \cite{optimumpreemption}, the optimum packet generation rates are obtained for self-preemptive and global preemptive bufferless systems for weighted AoI minimization, the latter case shown to allow closed-form expressions.

The most relevant existing analytical modeling work to this paper are the ones that study SBPSQ models for status update systems.  
The merits of SBPSQ systems are presented in \cite{pappas_etal_ICC15} in terms of lesser transmissions and AoI reduction.
The authors of \cite{moltafet_isit} derive the average AoI expressions for a two-source M/M/1/2 queueing system in which a packet waiting in the queue can be replaced only by a newly arriving packet from the same source using SHS techniques. The per-source MGF of the AoI is also obtained \cite{moltafet_wcomlet} for the two-source system by
using SHS under self-preemptive and non-preemptive policies, the latter being a per-source queueing system. 
However, in these works, the order of
packets in the queue does not change based on new arrivals and therefore AoI differentiation is not possible. 
\subsection{Scheduling Algorithms for Random Arrivals}
We now review the existing work on AoI scheduling with random arrivals that are related to the scope of the current paper. The authors of \cite{bedewy_etal_tit21} consider the problem of minimizing the age of information in a multi-source system and they show that for any given sampling strategy, the Maximum Age First (MAF) scheduling strategy provides the best age performance among all scheduling strategies. The authors of \cite{joo_eryilmaz_TNET18} propose an age-based scheduler that combines age
with the interarrival times of incoming packets, in its scheduling decisions, to achieve improved information freshness at
the receiver. Although the analytical
results are obtained for only heavy-traffic, their numerical results reveal that the proposed algorithm achieves desirable freshness performance for lighter loads as well.
The authors of \cite{kadota_tn18} and \cite{kadota_tmc21} consider an asymmetric  (source weights/service times are different) discrete-time wireless network with a base station serving multiple traffic streams using per-source queueing under the assumption of synchronized and random information packet arrivals, respectively, and propose nearly optimal age-based schedulers and age-agnostic randomized schedulers.
For the particular vase of random arrivals which is more relevant to the current paper, the reference \cite{kadota_tmc21} proposes a non-work-conserving stationary randomized policy for the single-buffer case with optimal scheduling probabilities depending on the source weights and source success probabilities through a square-root relationship and this policy is independent of the arrival rates. Moreover, they propose a work-conserving age-based Max-Weight scheduler for the same system whose performance is better and is close to the lower bound. We also note that similar results had been obtained in \cite{kadota_tn18} for synchronized arrivals. Our focus in this paper is on work-conserving age-agnostic schedulers that are more suitable for resource-constrained environments and  multi-hop scenarios for which it is relatively difficult to keep track of per-source AoI information at the server.

\section{Probabilistic Scheduling}
\label{section3} 
\subsection{Definitions of AoI and PAoI}
\begin{figure}[t]
	\centering
	\begin{tikzpicture}[scale=0.45]
	%\draw[step=1cm,gray,ultra thin,dotted] (0,0) grid (25,14);
	\draw[thick,<->,gray] (9,13) -- (16,13);
	\filldraw (12.5,13) circle (0.01) node[anchor=south, thick] {cycle-$j$};
	\draw[ultra thick,->] (0,0) -- (23,0) node[anchor=north] {$t$};
	\draw[ultra thick,->] (0,0) -- (0,14) node[anchor=west] {$\Delta^{(i)}(t)$};
	\draw[ultra thick,red] (4.5,4.5) -- (9,9);
	\filldraw[red] (4,4) circle (3pt);
	\filldraw[red] (3.5,3.5) circle (3pt) ;
	\filldraw[red] (3,3) circle (3pt); 
	\draw (0,9) node[anchor=east] {$\Phi_{j+1}^{(i)}$};
	\draw (0,5) node[anchor=east] {$\Psi_j^{(i)}$};
	\draw[dashed,gray] (0,9) -- (23,9);
	\draw[dashed,gray] (0,5) -- (23,5);
	\draw[dashed,gray] (0,2) -- (23,2);
	\draw[dashed,gray] (9,12.5) -- (9,0)  node[anchor=north, thick, black] {$T_{j}^{(i)}$};
	\draw[dashed,very thick,gray] (9,5) -- (4,0)  node[anchor=north, thick, black] {$A_{j}^{(i)}$};
	\draw[dashed,very thick, gray] (16,2) -- (14,0)  node[anchor=north east, thick, black] {$\quad A_{j+1}^{(i)}$};
	\draw[ultra thick,red] (9,9) -- (9,5.2);
	\draw[ultra thick,red] (9.1,5.1) -- (16,12);
	\draw (0,12) node[anchor=east] {$\Phi_{j}^{(i)}$};
	\draw (0,2) node[anchor=east] {$\Psi_{j+1}^{(i)}$};
	\draw[dashed,gray] (16,12.5) -- (16,0)  node[anchor=north, thick, black] {$T_{j+1}^{(i)}$};
%	\draw[dotted, red] (14,14) -- (14,0)  node[anchor=north,gray] {\tiny{14}};
%	\draw[dotted, red] (4,14) -- (4,0)   node[anchor=north,gray] {\tiny{4}};
%	\draw[dotted, gray] (23,14.5) -- (23,0)  node[anchor=north, gray] {\tiny{23}};
	
	\draw[dashed,gray] (0,12) -- (23,12);
	\draw[ultra thick,red] (16,12) -- (16,2.2);
	\draw[ultra thick,red] (16.1,2.1) -- (20,6);
	\filldraw[red] (20.5,6.5) circle (3pt);
	\filldraw[red] (21,7) circle (3pt) ;
	\filldraw[red] (21.5,7.5) circle (3pt); 
%	\draw (0,11.6) node[anchor=south west,gray] {\tiny{12}};
%	\draw (0,8.6) node[anchor=south west,gray] {\tiny{9}};
%	\draw (0,4.6) node[anchor=south west,gray] {\tiny{5}};
%	\draw (0,1.6) node[anchor=south west,gray] {\tiny{2}};
	\draw[red] (9,5) circle (6pt);
	\draw[red] (16,2) circle (6pt);
	\end{tikzpicture}
\caption{Sample path of the AoI process $\Delta^{(i)}(t)$.} 
\label{fig:samplepath}
\end{figure}
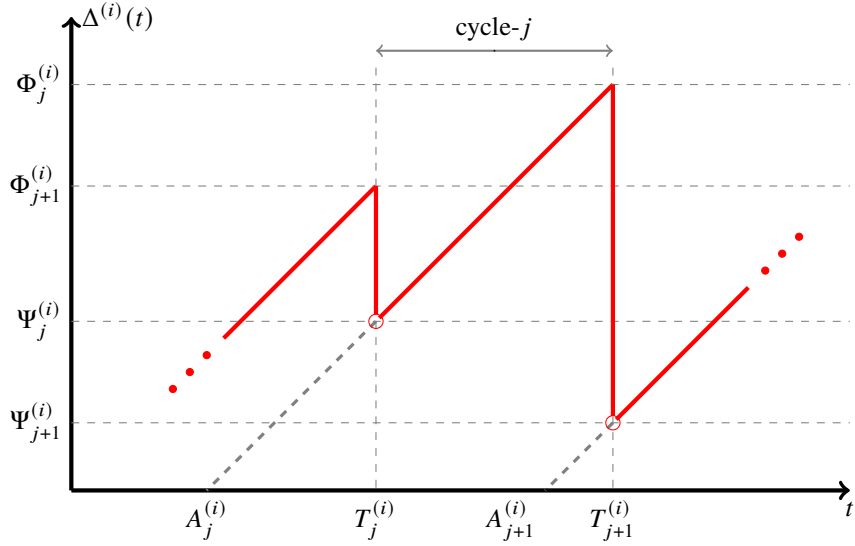

In a very general setting, let $T_j^{(i)}$ and $A_j^{(i)}$ for $j\geq 1$ denote the times
at which the $j$th successful source-$i$ packet is received by the monitor and generated at the source, respectively. We also let $\Psi^{(i)}_j$ denote the system time of the $j$th successful source-$i$ information packet which is the sum of the packet's queue wait time and service times, i.e., $\Psi^{(i)}_j=T_j^{(i)} - A_j^{(i)}$. Fig.~\ref{fig:samplepath} depicts a sample path of the source-$i$ AoI process $\Delta^{(i)}(t)$ which increases with unit slope from the value $\Phi_{j}^{(i)}$ at $t=T_{j}^{(i)}$ until $t=T_{j+1}^{(i)}$ in cycle-$j$. The peak value in cycle-$j$ is denoted by $\Psi_{j}^{(i)}$ which represents the Peak AoI process for source-$i$. These definitions apply to general status update systems. Note that for the specific system of Fig.~\ref{fig:twosource}, successful packets are the ones which are received by the monitor, and those that are replaced by fresher incoming packets while at the waiting room are unsuccessful packets. Let $\Delta^{(i)}$ and $\Phi^{(i)}$ denote the steady-state values for the source-$i$ processes $\Delta^{(i)}(t)$ and $\Phi_j^{(i)}$, respectively. The weighted average AoI, $W_{AoI}$, and the weighted average PAoI, $W_{PAoI}$, of the system are written as
\begin{equation}
W_{AoI} = \sum_{i=1}^2 \omega_i E [ \Delta^{(i)}], 
\ W_{PAoI}= \sum_{i=1}^2 \omega_i  E [ \Phi^{(i)}], \label{W}
\end{equation} where $\omega_i, i=1,2,$ with $\omega_1+\omega_2=1$ are the (normalized) weighting coefficients.
\subsection{System Model}
In this paper, we consider a non-preemptive status update system in Fig.~\ref{fig:twosource} with two sources, a server, and a monitor. Source-$i$, $i=1,2$ generates information packets (containing time-stamped status update information) according to a Poisson process with intensity $\lambda_i$. The generated packets become immediately available at the server.
The server maintains two single-buffer queues, namely $Q_i, i=1,2$, that holds the freshest packet from source-$i$. This buffer management is referred to as Single-Buffer Per-Source Queueing (SBPSQ). A newcoming source-$i$ packet receives immediate service if the server is idle and there are no waiting packets, or joins the empty $Q_i$, or replaces the existing staler source-$i$ packet at $Q_i$. The server is work-conserving as a result of which an information packet is immediately transmitted unless the system is idle. Consequently, when the system has one packet waiting at $Q_i$ for $i=1$ or $i=2$ upon the server becoming idle, then this packet from $Q_i$ will immediately be served. When there are two packets waiting at the two queues, then the server is to transmit the packet from $Q_i$ with probability $p_i$ with $p_1 + p_2 =1$. Therefore, the scheduler is age-agnostic and does not require the server to read the timestamp field in the packets and keep track of the instantaneous AoI values. The probabilities $p_i$'s are to be chosen so as to provide AoI/PAoI differentiation. 
%In this paper, we attempt to provide differentiation through the minimization of a certain weighted average AoI/PAoI. The motivation behind AoI differentiation is that in  a networked control system, the information about certain input stochastic processes need to be kept relatively fresher at the control unit since this information will have profound impact on the performance of the system.
At the end of a single transmission, positive/negative acknowledgments from the monitor to the server are assumed to be immediate, for the sake of convenience. The channel success probability for source-$i$ is $s_i$ and when a packet's transmission gets to start, it will be retransmitted until it is successfully received by the monitor.
Therefore, if a single transmission is assumed to be exponentially distributed with parameter $\nu_i$, then the transmission time of successful source-$i$ packets from the server to the monitor are exponentially distributed with parameter $\mu_i = \nu_i s_i$ by taking into account of the retransmissions.  With this choice, error-prone channels are also considered in this paper.
We define the source-$i$ load as $\rho_i = \frac{\lambda_i}{\mu_i}$ and the total load $\rho = \rho_1+\rho_2$.
We also define the traffic mix parameter $r_i$ so that $\rho_i =\rho r_i$, and the traffic mix ratio $r=\frac{r_1}{r_2}$.
The studied model falls in the general framework of status update systems analytically studied in the literature; see the surveys on AoI \cite{kosta_etal_survey},\cite{survey_Yates} and the references therein for a collection of multi-source queueing models for AoI.

The analytical method we propose in the next subsection enables us to obtain the distribution of $\Delta^{(1)}$ and $\Phi^{(1)}$. By renumbering the sources, the distribution of $\Delta^{(2)}$ and $\Phi^{(2)}$ can also be obtained using the same method. 
\subsection{Queueing Model}
\begin{table}[tb]
	\begin{center}
		\begin{tabular}{||c|c|c|c||} 
			\hline
			State & Server & $Q_1$ & $Q_2$ \\ [0.5ex] 
			\hline\hline
			1 & I & E & E \\ 
			\hline
			2 & B1 & E & E \\
			\hline
			3 & B1 & F & E \\
			\hline
			4 & B1 & E & F \\
			\hline
			5 & B1 & F & F \\  
			\hline
			6 & B2 & E & E \\
			\hline
			7 & B2 & F & E \\
			\hline
			8 & B2 & E & F \\
			\hline
			9 & B2 & F & F \\ [0.1ex]
			\hline \hline
		\end{tabular}
	\end{center}
	\caption{Description of the 9 states of the CTMC $\bm{X}(t)$. I, E, and F, stand for idle, empty, and full, respectively. B1 (B2) stands for the server being busy serving a source-1 (source-2) packet.}
	\label{step1}
\end{table}
The proposed method consists of two main steps. In the first step, we construct an irreducible Continuous Time Markov Chain (CTMC) denoted by $\bm{X}(t)$ with nine states each of which is described in detail in Table~\ref{step1}. The CTMC $\bm{X}(t)$ has the generator matrix $\bm{P}$ where
\begin{align}
\bm{P_0} & = \begin{pmatrix}
0 & \lambda_1 & 0 & 0 & 0 & \lambda_2 & 0 & 0 & 0 \\
\mu_1 & 0 & 0 & \lambda_1& \lambda_2 & 0 & 0 & 0 & 0 \\
0 & \mu_1 & 0 & 0 & 0 & \lambda_2 &  0 & 0 & 0  \\
0 & 0 & 0 & 0 &  \lambda_1 & \mu_1 & 0 & 0 & 0 \\
0 & 0 & 0 & \mu_1 p_1 & 0 & 0 & \mu_1 p_2 & 0 & 0 \\
\mu_2 & 0 & 0 & 0 & 0 & 0  &\lambda_1 & \lambda_2 &0\\
0&  \mu_2 & 0 & 0 & 0 & 0 & 0 & 0 & \lambda_2 \\
0& 0 & 0 & 0 &  0 &  \mu_2 & 0 & 0 & \lambda_1 \\
0& 0 & 0 & \mu_2 p_1 & 0 & 0 & \mu_2 p_2 & 0 & 0
\end{pmatrix},
\end{align}
and $\bm{P}$ is the same as $\bm{P_0}$ except for its diagonal entries which are set to the corresponding row sums with a minus sign so that $\bm{P} \bm{1} =\bm{0}$ where $\bm{1}$ and $\bm{0}$ are column vectors of ones and zeros, respectively, of appropriate size. Let $\bm{\pi}$ be the stationary solution for $\bm{X}(t)$ so that
\begin{align}
\bm{\pi} \bm{P} = 0, \ \bm{\pi} \bm{1}=1, 
\end{align}
with $\bm{\pi_j}$ denoting the steady-state probability of any new packet arrival finding the system in state $j$.
%\subsection{Subsection Heading Here}
%Subsection text here.

In the second step of the proposed method, we construct an absorbing CTMC denoted by $\bm{Y}(t)$ with 14 transient states $1,2,\ldots,14$ and two absorbing states $15,16$ which starts to evolve with the arrival of a source-1 packet, say packet $n$ into the system. If this packet turns out to be unsuccessful then we transition to the absorbing state 15. If packet $n$ turns out to be successful, then we evolve until the reception of the next successful packet say $m$ at which point the absorbing state 16 is transitioned to, which is referred to as a successful absorption. The 14 transient states are described in Table~\ref{step2}.
\begin{table}[tb]
	\begin{center}
		\begin{tabular}{||c|c|c|c|c||} 
			\hline
			State & Server & Packet $n$ & $Q_1$ & $Q_2$ \\ 
		 [0.5ex] 
			\hline\hline
			1 & \bf{B1}& N& E & E \\ 
			\hline
			2 & \bf{B1} & N & E  & E \\
			\hline
			3 & B1 & N & F & E \\
			\hline
			4 & \bf{B1}  & N & E & F \\
			\hline
			5 & \bf{B1}  & N &  F & F \\  
			\hline
			6 & B1  & N &  F & F \\  
			\hline
			7 & B2 & N & E & F \\
			\hline
			8  & B2 & N & F & F \\
			\hline
			9 & I & Y & E & E \\
			\hline
			10 & B1  & Y & X & X \\ \hline
			11 & B2 & Y & E & E \\ \hline
			12 & B2 & Y & F & E \\ \hline 
			13 &  B2 & Y & E & F \\ \hline 
			14 &  B2 & Y & F & F \\ 
			 [0.1ex]
			\hline \hline
		\end{tabular}
	\end{center}
	\caption{Description of the 14 states of the CTMC $\bm{Y}(t)$. I, E, and F, stand for idle, empty, and full, respectively, and B1 (B2) stands for the server being busy serving a source-1 (source-2) packet. The notation {\bf B1} means the particular packet $n$ is being served. N and Y stand for packet $n$ not successful yet and otherwise, respectively, and X is don't care.}
	\label{step2}
\end{table}
The generator for the absorbing CTMC, denoted by $\bm{Q}$ is in the form
\begin{align}
\bm{Q}=\begin{pmatrix}
\bm{A} & \bm{u} & \bm{s}  \\
\bm{0} & 0 & 0 \\
\bm{0} & 0 & 0 
\end{pmatrix},
\end{align}
where 
	\begin{align}
	\bm{A_0} & = 
	\left(
	\begin{array}{cccccccccccccc}
	0 & \lambda_1 & 0 & \lambda_2 & 0 & 0 & 0 & 0 & \mu_1 & 0 & 0 & 0 & 0 &  0 \\
	0 & 0 & 0 & 0 & \lambda_2 & 0 &  0 & 0 & 0 & \mu_1 & 0 &  0 & 0 & 0  \\
	\mu_1 & 0 & 0 & 0 & 0 &  \lambda_2 & 0 & 0 & 0 & 0 & 0 & 0 & 0 &  0 \\
	0 & 0 & 0 & 0 & \lambda_1 & 0 & 0 & 0 & 0 & 0 & \mu_1 & 0 & 0 & 0 \\
	0 & 0 & 0 & 0 & 0 & 0 & 0 & 0 & 0 & \mu_1 p_1 & 0 & \mu_1 p_2 & 0 & 0 \\
	0 & 0 & 0 &  \mu_1 p_1 & 0 & 0 & \mu_1 p_2 & 0 & 0 & 0 & 0 & 0 & 0 & 0 \\
	\mu_2 & 0 &  0 & 0 & 0 & 0 & 0 & \lambda_2 & 0 & 0 & 0 & 0 & 0 & 0 \\ 
	0 & 0 & 0 & \mu_2 p_1 & 0 & 0 & \mu_2 p_2 & 0 & 0 & 0 & 0 & 0 & 0 & 0 \\
	0 & 0 & 0 & 0 & 0 & 0 & 0 & 0 & 0 & \lambda_1 & \lambda_2 & 0 & 0 & 0 \\ 
	0 & 0 & 0 & 0 & 0 & 0 & 0 & 0 & 0 & 0 & 0 & 0 & 0 & 0 \\
	0 & 0 & 0 & 0 & 0 & 0 & 0 & 0 & \mu_2 &0 &0 & \lambda_1 & \lambda_2 & 0 \\
	0 & 0 & 0 & 0 & 0 & 0 & 0 & 0 & 0 & \mu_2 & 0 &  0 & 0 & \lambda_2 \\
	0 & 0 & 0 & 0 & 0 & 0 & 0 & 0 & 0 & 0 & \mu_2 & 0 & 0 & \lambda_1 \\
	0 & 0 & 0 & 0 & 0 & 0 & 0 & 0 & 0 & \mu_2 p_1 & 0 & \mu_2 p_2 & 0 & 0 \\
	\end{array}
	\right),
	\bm{u} = \begin{pmatrix}
	0 \\
	0  \\
	\lambda_1 \\
	0 \\
	0 \\
	\lambda_1 \\
	\lambda_1 \\
	\lambda_1 \\
	0 \\
	0 \\
	0 \\
	0 \\
	0 \\
	0
	\end{pmatrix},
	\bm{s} = \begin{pmatrix}
	0 \\
	0  \\
	0 \\
	0 \\
	0 \\
	0 \\
	0 \\
	0 \\
	0 \\
	\mu_1 \\
	0 \\
	0 \\
	0 \\
	0
	\end{pmatrix},
	\bm{h} = \begin{pmatrix}
	0 \\
	0  \\
	0 \\
	0 \\
	0 \\
	0 \\
	0 \\
	0 \\
	1 \\
	1 \\
	1 \\
	1 \\
	1 \\
	1
	\end{pmatrix}
	\label{big}
	\end{align}
 $\bm{A}$ is the same as $\bm{A_0}$ in \eqref{big} except for its diagonal entries which are set to the corresponding row sums with a minus sign so that $\bm{A} \bm{1} + \bm{u} + \bm{s}=\bm{0}$. Note that $\bm{A}$ is the sub-generator matrix corresponding to the transient states and $\bm{u}$ and $\bm{s}$
are the transition rate vectors from the transient states to the unsuccessful and successful absorbing states, respectively. The vector $\bm{h}$ which takes the unit value for the indices 9 to 14, and zero otherwise, will be needed in deriving the AoI distribution. 

The initial probability vector of the CTMC  $\bm{Y}(t)$ is denoted by $\bm{\alpha}$ which is given as follows:
\begin{align}
\bm{\alpha} & = \begin{pmatrix}
\bm{\pi_1} & 0 & \bm{\pi_{23}}  & \bm{0_{1 \times 2}} & \bm{\pi_{45}} & \bm{\pi_{67}} & \bm{\pi_{89}} & \bm{0_{1 \times 6}}
\end{pmatrix},
\end{align}
where $\bm{\pi_{ij}} := \bm{\pi_i} + \bm{\pi_j}$. In order to understand this, a new source-1 packet $n$ will find the system idle (state 1 of ${\bm X(t)}$) with probability $\bm{\pi_1}$ and therefore will be placed in service immediately, i.e., state 1 of ${\bm Y(t)}$. Similarly, packet $n$ will find the system in states 2 and 3 of  ${\bm X(t)}$ with probability $\bm{\pi_{23}}$ and in either case this packet will start its journey from state 3 of ${\bm Y(t)}$ and so on. With this step, the two CTMCs ${\bm X(t)}$ and ${\bm Y(t)}$ are linked.

Let us visit Fig.~\ref{fig:samplepath} and relate it to the absorbing CTMC ${\bm Y(t)}$. The instance $A_{j}^{(i)}$ is the arrival time of packet $n$ of  ${\bm Y(t)}$  and $T_{j+1}^{(i)}$ is the reception time of packet $m$. Therefore, the distribution of the absorption times of  ${\bm Y(t)}$ in successful absorptions enables us to write the steady-state distribution of the PAoI process. In particular,
\begin{align}
\Pr \{ \Phi^{(1)} \leq x \} & = \Pr \{ {\bm Y(x)} =16 \ | \ {\bm Y(\infty)} = 16\} \\
& = \frac{\Pr \{ {\bm Y(x)} =16 \}}{\Pr\{ {\bm Y(\infty)} = 16 \} } 
% -\bm{\alpha} \bm{A^{-1}} \bm{s}
\end{align}
Differentiating this expression with respect to $x$, we obtain the pdf (probability density function) of  $\Phi^{(1)}$, denoted by $f_{ \Phi^{(1)}}(x)$, as follows;
\begin{align}
f_{\Phi^{(1)}}(x)  & = \beta \ \bm{\alpha} \mathrm{e}^{\bm{A}x} \bm{s},
\end{align}
where $\beta^{-1} =\Pr\{ {\bm Y(\infty)} = 16 \} = -\bm{\alpha} \bm{A^{-1}} \bm{s} $. 

Revisiting Fig.~\ref{fig:samplepath}, the probability $\Pr \{ x < \Delta^{(1)} \leq  x + \delta x \}$ is proportional with $\Pr \{  {\bm Y(x)} \in \cal{S}  \}$
with the subset $\cal{S}$ containing the six transient states 9 to 14 of ${\bm Y(t)}$  and the proportionality constant being the reciprocal of the mean holding time in  $\cal{S}$ in successful absorptions.
Consequently, we write
\begin{align}
f_{ \Delta^{(1)}}(x) & = \kappa \ \bm{\alpha} \mathrm{e}^{\bm{A}x} \bm{h},
\end{align}
where $\kappa^{-1} = -\bm{\alpha} \bm{A^{-1}} \bm{h} $. The $k$th non-central moments of $\Phi^{(1)}$ and $\Delta^{(1)}$ are subsequently very easy to write:
\begin{align}
E \left[(\Phi^{(1)})^k\right] & =  \beta \  \bm{\alpha}( -\bm{A})^{-k-1} \bm{s}, \quad  
E \left[ (\Delta^{(1)})^k \right]  =  \kappa \ \bm{\alpha}( -\bm{A})^{-k-1} \bm{h}. \label{momentsPA}
\end{align}
\section{Heavy-traffic Regime}
\label{section4}
In this section, we study the so-called heavy-traffic  regime, i.e., $\lambda_i \rightarrow \infty$. We first describe the analytical model in this regime along with the closed-form average AoI/PAoI expressions. Subsequently, we propose two heuristic schedulers based on this model that are devised to operate at any load as well as an optimum probabilistic scheduler on the basis of the analytical model of the previous section. 
\subsection{Analytical Model}
In this case, the CTMC in step 1 of the proposed method reduces to one single state corresponding to a busy server with both queues being full since in the heavy-traffic regime, neither the queues can be empty nor the server can be idle. Moreover, the absorbing CTMC with 14 transient and 2 absorbing states reduces to one with 3 transient states and 1 successful absorbing state (the state $\bm{s}$). The transient states 1 and 3 indicate that packet $n$ and packet $m$ are in service, respectively, whereas transient state 2 indicates the transmission of a source-2 packet. Consequently, the matrices characterizing this absorbing CTMC take the following simpler form:
	\begin{align}
	\bm{A} & = 
	\left(
	\begin{array}{ccc}
	-\mu_1 & \mu_1 p_2 & \mu_1 p_1 \\
	0 & -\mu_2 p_1 & \mu_2 p_1 \\
	0 & 0 & -\mu_1
	\end{array}
	\right),
	\bm{s} = \begin{pmatrix}
	0 \\
	0 \\
	\mu_1
	\end{pmatrix},
	\bm{h} = \begin{pmatrix}
	0 \\
	1 \\
	1 
	\end{pmatrix}, {\bm \alpha}=\begin{pmatrix}
	1 \\ 
	0 \\
	0 
	\end{pmatrix}^T,
	\label{small}
	\end{align}
and the expressions in \eqref{momentsPA} are valid for the moments of AoI/PAoI in this heavy-traffic regime. Using the upper-triangular nature of the matrix $\bm{A}$ and \eqref{momentsPA}, it is not difficult to show that
 \begin{align}
 E [ \Phi^{(1)}] & = \frac{2}{\mu_1} + \frac{p_2}{\mu_2 p_1}, \ E [ \Phi^{(2)}] = \frac{2}{\mu_2} + \frac{p_1}{\mu_1 p_2}. \label{nail11}
 \end{align} 
Defining the probability ratio $p=\frac{p_1}{p_2}$ and the weight ratio $\omega=\frac{\omega_1}{\omega_2}$, the weighted average PAoI simplifies to 
\begin{align}
W_{PAoI} & = \frac{\omega_2}{\mu_1 \mu_2} \left(  2 \omega (\mu_1 + \mu_2) + \omega \mu_1 p^{-1} + \mu_2 p \right)
\end{align}
Employing the  Karush–Kuhn–Tucker (KKT) conditions on this expression and defining $\mu = \frac{\mu_1}{\mu_2}$, the optimum probability ratio that yields the minimum $W_{PAoI}$, denoted by $p_{PAoI}^*$, can easily be shown to satisfy the following:
\begin{align}
p_{PAoI}^* &  \; {\propto } \; \sqrt{\omega \mu}. \label{HL}
\end{align}
The expression for the average AoI is somewhat more involved:
\begin{align}
E [ \Delta^{(1)}] & = \frac{1}{\mu_1} + \frac{\mu_2 p_1 + \mu_1}{\mu_1 \mu_2 p_1} - \frac{1}{\mu_2 p_1 + \mu_1 p_2}. \label{nail10}
\end{align} 
A similar expression for $E [ \Delta^{(2)}]$ is easy to write due to symmetry. However, in this case, the KKT conditions for the expression for $W_{AoI}$ give rise to a quartic equation, i.e, 4th degree polynomial equation, for the roots of which closed-form expressions are not available. However, numerical techniques can be used to find the optimum probability ratio minimizing $W_{AoI}$, denoted by $p_{AoI}^*$, in this case. 
However, for the special case $\mu_1 = \mu_2 = u$, the expression \eqref{nail10} reduces to
\begin{align}
E [ \Delta^{(1)}] & = \frac{1}{u} + \frac{1}{u p_1}, \ E [ \Delta^{(2)}]  = \frac{1}{u} + \frac{1}{u p_2},
\end{align} 
which are identical to the expressions for $E [ \Phi^{(1)}]$ and $E [ \Phi^{(2)}]$ in \eqref{nail11}, respectively, for the special case  
$\mu_1 = \mu_2 = u$.
Employing KKT conditions on $W_{AoI}$, it is obvious to show that
\begin{align}
p_{AoI}^* &  \; {\propto } \; \sqrt{\omega}. \label{HL2}
\end{align}
When $\mu_1 \neq \mu_2$, we use exhaustive search to obtain $p_{AoI}^*$ throughout the numerical examples of this paper.
\subsection{Proposed Heuristic Schedulers}
The focus of this paper is on work-conserving schedulers that are neither age- or timestamp-aware, i.e., the schedulers make a decision only on the source indices of packets in the waiting room, and not on the timestamp information in the packets or the instantaneous ages of the source processes.
This allows us to use simple-to-implement scheduling policies without the server having to  process the timestamp information included in the information packets.

Given the traffic parameters $\lambda_i, \mu_i,$ and the weights $\omega_i$, for $i=1,2$, we first introduce the OPS-P (Optimum Probabilistic Scheduling for PAoI) policy that minimizes the weighted average PAoI of the system given in \eqref{W}.
OPS-A (Optimum Probabilistic Scheduling for PAoI) is defined similarly so as to minimize the weighted average AoI in \eqref{W}.
We use the analytical model and exhaustive search to obtain OPS-P and OPS-A. Although the analytical model is computationally efficient, one needs to resort to simpler heuristics which may be beneficial especially in situations where the traffic parameters may vary in time and the server may need to update its scheduling policy without having to perform extensive computations.  
For this purpose, we propose a generic heuristic probabilistic scheduler called H1$(p)$ that employs the probability ratio $p=\frac{p_1}{p_2}, p_1=\frac{p}{1+p}, p_2=\frac{1}{1+p},$ using the information about $\omega$ and $\mu$ only but not the actual arrival rates $\lambda_i, i=1,2$.
The second heuristic scheduler we propose is called H2$(p)$ which is obtained by determinizing the probabilistic policy H1$(p)$ as described below. In H2$(p)$, each source-$i$ maintains a bucket $b_i$ so that $b_1 + b_2=0$ at all times. Initially, $b_i=0, i=1,2$. When there are two packets in the waiting room, the source with the larger  bucket value $b_i$ is selected for transmission. Every time a source-$1$ packet is transmitted, $b_1$ is decremented by $(1 - p_1)$ and $b_2$ is incremented by $p_2$.  Similarly, when a source-$2$ packet is transmitted, $b_2$ is decremented by $(1 - p_2)$ and $b_1$ is incremented by $p_1$. 
In order for the bucket values not to grow to infinity (which may occur if there are no packet arrivals from a specific source for an extended duration of time), we impose a limit on the absolute values of the buckets, i.e., $| b_i | < B$ where $B$ is called the bucket limit. 

Note that in the heavy-traffic regime, H2$(p)$ is the determinized version of H1$(p)$. To see this, let $p=1,p_1=p_2=0.5$. In H1$(p)$, a geometrically distributed (with parameter 0.5) number of source-1 packets will be transmitted followed with the transmission of a geometrically distributed (again with parameter 0.5) number of source-2 packets. On the other hand, in H2$(1)$, an alternating pattern arises where a single source-1 packet transmission is to be followed by a single packet-2 transmission, i.e., round-robin scheduling. For both heuristic schedulers, the ratio of source-1 transmissions to source-2 transmissions is kept at $p$ in the heavy-traffic regime, but H2$(p)$ manages to maintain this ratio using deterministic patterns as opposed to being probabilistic. The bucket-based nature of the algorithm enables one to obtain this deterministic pattern for all values of the ratio parameter $p$ which is advantageous especially for average AoI. 
Moreover, in H2$(p)$, we seek to maintain a probability ratio $p$ of transmissions between the two sources throughout the entire operation of the system whereas this probability ratio is maintained in H1$(p)$ only during times when there are two packets in the waiting room.
When we choose $p=p_{PAoI}^*$ with $p_{PAoI}^*$ being the optimum probability ratio in the heavy-traffic regime (see Eqn.~\eqref{HL}), we obtain our two proposed schedulers H1-P (Heuristic 1 Scheduler for PAoI) and H2-P (Heuristic Scheduler 2 for PAoI) for average weighted PAoI minimization, i.e., 
\begin{align}
\text{H1-P} & \equiv \text{H1}(p_{PAoI}^*), \ \text{H2-P} \equiv \text{H2}(p_{PAoI}^*),
\end{align}	 where the notation $\equiv$ is used to denote equivalence.
Similarly, we propose two schedulers for average weighted AoI minimization, namely H1-A (Heuristic 1 Scheduler for AoI) and H2-A (Heuristic Scheduler 2 for AoI) , i.e., 
\begin{align}
\text{H1-A} &\equiv \text{H1}(p_{AoI}^*),
 \ 	 \text{H2-A} \equiv  \text{H2}(p_{AoI}^*).
\end{align} For two-source networks with $\mu=1$, H1-P $\equiv$ H1-A, and  H2-P $\equiv$ H2-A.
%The results for the non-preemptive bufferless NPB system are also obtained as a benchmark.  

\section{Analytical Model for the Non-preemptive Bufferless Server}
\label{section5}
Up to now, we have considered SBPSQ servers with scheduling. In this section, we also study the Non-Preemptive Bufferless (NPB) server for the purpose of using it as a benchmark against the per-source queueing  systems of our interest. 
In the NPB scenario, the newcoming source-$i$ packet is served immediately if the server is idle or is otherwise discarded since there is no waiting room. 
Actually, an analytical model for AoI and PAoI is recently proposed in \cite{dogan_akar_tcom21} for servers serving a general number of sources with more general phase-type distributed service times, also allowing arbitrary preemption probabilities. In this section, we make use of the model introduced by \cite{dogan_akar_tcom21} to provide closed-form expressions for the average AoI/PAoI for the specific case of two sources, no preemption, and exponentially distributed service times. While doing so, we use absorbing CTMCs as opposed to Markov Fluid Queues (MFQ) used in \cite{dogan_akar_tcom21}. Both yield the same results but ordinary CTMCs of absorbing type are more commonly known and established than MFQs.
In this case, the CTMC in step 1 is not needed due to the bufferless nature of the system. 
 Moreover, the absorbing CTMC with 14 transient and 2 absorbing states reduces to one with 4 transient states and 1 absorbing state. The transient states 1 and 4 indicate that packet $n$ and packet $m$ are in service, respectively, whereas in transient state 2, we wait for a packet arrival, and in transient state 3, a source-2 packet is in service. Consequently, the matrices characterizing this absorbing CTMC are written as:
\begin{align}
\bm{A} & = 
\left(
\begin{array}{cccc}
-\mu_1 & \mu_1  & 0 & 0 \\
0 & -(\lambda_1+\lambda_2) & \lambda_2 & \lambda_1  \\
0 & \mu_2 & -\mu_2 & 0 \\
0 & 0  & 0  & -\mu_1
\end{array}
\right), \
\bm{s} = \begin{pmatrix}
0 \\
0 \\
0 \\
\mu_1
\end{pmatrix},\
\bm{h} = \begin{pmatrix}
0 \\
1 \\
1 \\
1
\end{pmatrix}, \ 
{\bm \alpha}  =\begin{pmatrix}
1 \\
0 \\
0 \\
0
\end{pmatrix}^T,
\label{bufferless}
\end{align}
and the expressions \eqref{momentsPA}  can be used for obtaining the moments of AoI/PAoI for the bufferless system. Using \eqref{momentsPA}, for the average per-source PAoI, one can easily show that
\begin{align}
E [ \Phi^{(1)}] & = \frac{1}{\mu_1} + \frac{(1+\rho)}{\lambda_1}, \ E [ \Phi^{(2)}] = \frac{1}{\mu_2} + \frac{(1+\rho)}{\lambda_2}.
\end{align} 
Recalling the definition of the traffic mix parameter $r_i$ and the traffic mix ratio $r=\frac{r_1}{r_2}$, the weighted average PAoI can be written in terms of $r_1$ as follows: 
\begin{align}
W_{PAoI} & = \frac{\omega_1}{\mu_1} + \frac{\omega_2}{\mu_2} + \frac{\omega_1(1+\rho)}{\rho r_1 \mu_1} + \frac{\omega_2(1+\rho)}{\rho (1-r_1) \mu_2}.
\end{align}
Fixing $\rho$ and employing the KKT conditions for this expression, the optimum traffic mix ratio, denoted by $r_{PAoI}^*$, is given as:
\begin{align}
r_{PAoI}^* &  \; {\propto } \; \sqrt{\frac{\omega}{\mu}}. \label{optimum_mix_ratio}
\end{align}
Note that the above ratio does not depend on the load parameter $\rho$.
If we define the arrival rate ratio $\lambda = \frac{\lambda_1}{\lambda_2}$, then the optimum arrival rate ratio, denoted by $\lambda_{PAoI}^*$, can be written as:
\begin{align}
\lambda_{PAoI}^* &  \; {\propto } \; \sqrt{{\omega}{\mu}}. \label{optimum_arrivalrate_ratio}
\end{align}
The expression for the average AoI can be written as:
\begin{align}
E [ \Delta^{(1)}] & = \frac{1}{\mu_1 \mu_2} \left( {\mu_2} + \frac{\mu_2}{\rho_1} + \frac{\lambda_2}{\rho_1} +  \frac{\mu_2 \rho_1 + \mu_1 \rho_2}{(1+\rho)} \right) . 
\end{align} 
A similar expression for $E [ \Delta^{(2)}]$ is again very easy to write due to symmetry. However, in this case, the KKT conditions for the expression for $W_{AoI}$ again result in a quartic equation in which case numerical techniques can be used to find the optimum traffic mix ratio denoted by $r_{AoI}^*$.

\section{Numerical Examples}
\label{section6}
\subsection{Heavy-traffic Scenario}
\begin{figure}[bth]
	\centering
	\includegraphics[width=0.7\linewidth]{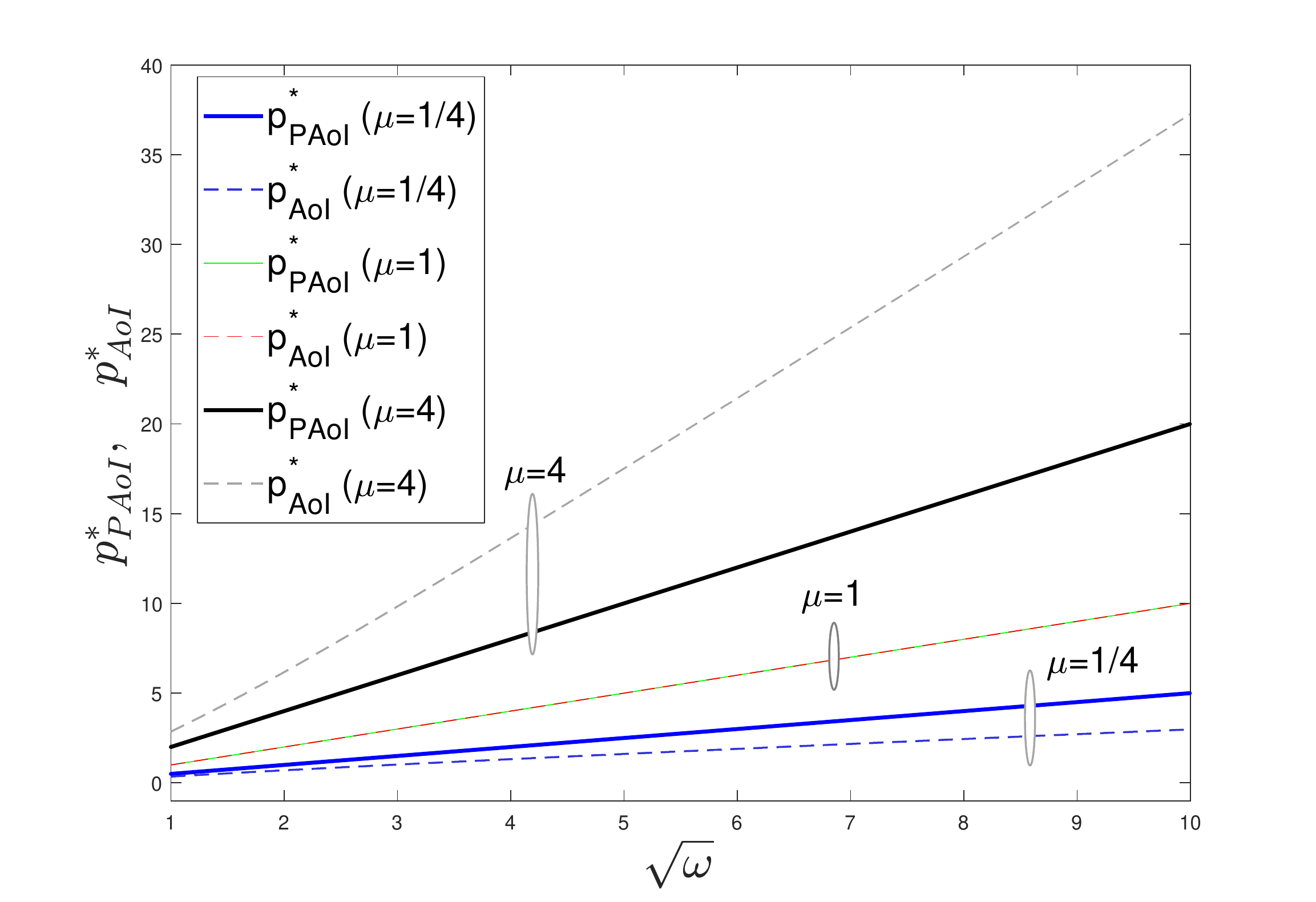}
	\caption{The probability ratio parameters $p_{PAoI}^*$ and $p_{AoI}^*$ as a function of the square root of the weight ratio parameter, $\sqrt{\omega}$, for three values of the service rate ratio parameter $\mu$. 
	} 
	\label{fig:ornek1}
\end{figure}

\begin{figure}[bth]
	\centering
	\includegraphics[width=0.7\linewidth]{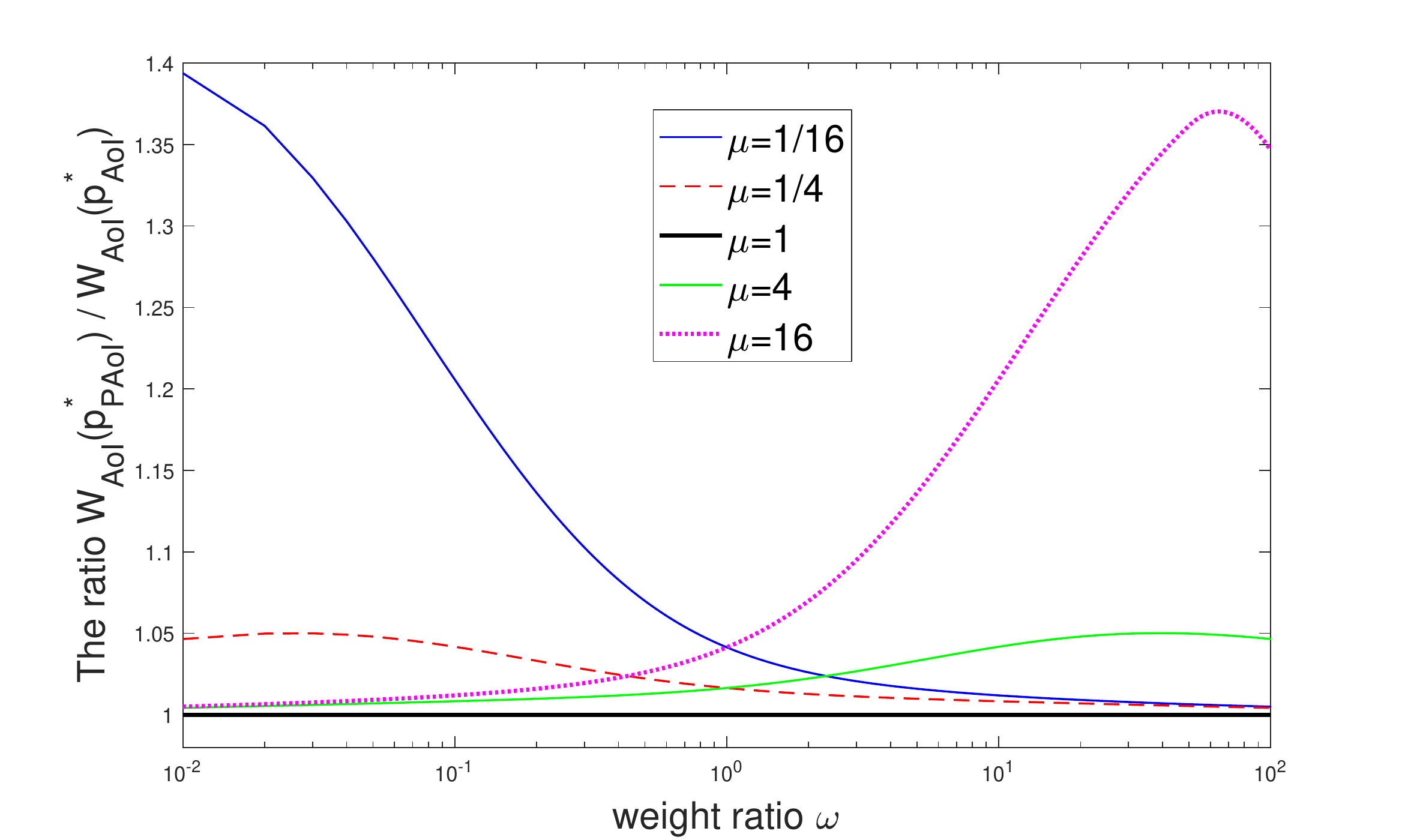}
	\caption{The ratio of $W_{AoI}$ obtained with the use of $p_{PAoI}^*$ to that using $p_{AoI}^*$ as a function of the weight ratio parameter, ${\omega}$, for five values of the service rate ratio parameter $\mu$. 
	} 
	\label{fig:ornek1b}
\end{figure}

In the first numerical example, we study the heavy-traffic regime and we depict the corresponding optimum probability ratio parameters $p_{PAoI}^*$ and $p_{AoI}^*$ as a function of the square root of the weight ratio parameter, $\sqrt{\omega}$, for three values of the service rate ratio parameter $\mu$ in Fig.~\ref{fig:ornek1}. When the service rates of the two sources are identical, then these probability ratios are the same for both AoI and PAoI. However, when the service rate ratio starts to deviate from unity, then the optimum probability ratio parameters for PAoI and AoI turn out to deviate from each other. More specifically, $p_{AoI}^* < p_{PAoI}^*$ when $\mu < 1$ and $p_{AoI}^* > p_{PAoI}^*$ when $\mu > 1$. Subsequently, we study whether one can use the easily obtainable $p_{PAoI}^*$ in Eqn.~\eqref{HL} in place of  $p_{AoI}^*$ when the minimization of weighted AoI is sought. 
Fig.~\ref{fig:ornek1b} depicts the ratio of $W_{AoI}$  obtained with the use of the probability ratio $p_{PAoI}^*$ to that obtained using $p_{AoI}^*$ as a function of the weight ratio $\omega$. We observe that $p_{PAoI}^*$ can be used in place of  $p_{AoI}^*$ only when the rate ratio $\mu$ and the weight ratio $w$ are both close to unity. It is clear that when $\mu=1$, the depicted ratio in Fig.~\ref{fig:ornek1b} is always one irrespective of $\omega$; also see \eqref{HL2}.
\subsection{Numerical Study of the Proposed Schedulers}
A two-source network is called symmetric when $\omega=1, \ \mu=1$ in \cite{kadota_tmc21} and is asymmetric otherwise. We first present our numerical results for symmetric networks and subsequently, asymmetric network results are presented first for weighted average PAoI minimation, and then for weighted average AoI minimization. We fix $\mu_2=1$ in all the numerical examples. Thus, one time unit is taken as the average service time of source-2 packets. All the results are obtained through the analytical models developed in this paper except for the bucket-based H2-P and H2-A for which an analytical model is cumbersome to build for all values of the probability parameter $p$ and therefore we resorted to simulations.
\subsection{Symmetric Network}
\begin{figure}
	\centering
	\begin{subfigure}[b]{0.48\textwidth}
		\centering
			\caption{$W_{PAoI}$ ($\rho=1$)} \vspace*{-0.3cm}
		\includegraphics[width=\textwidth]{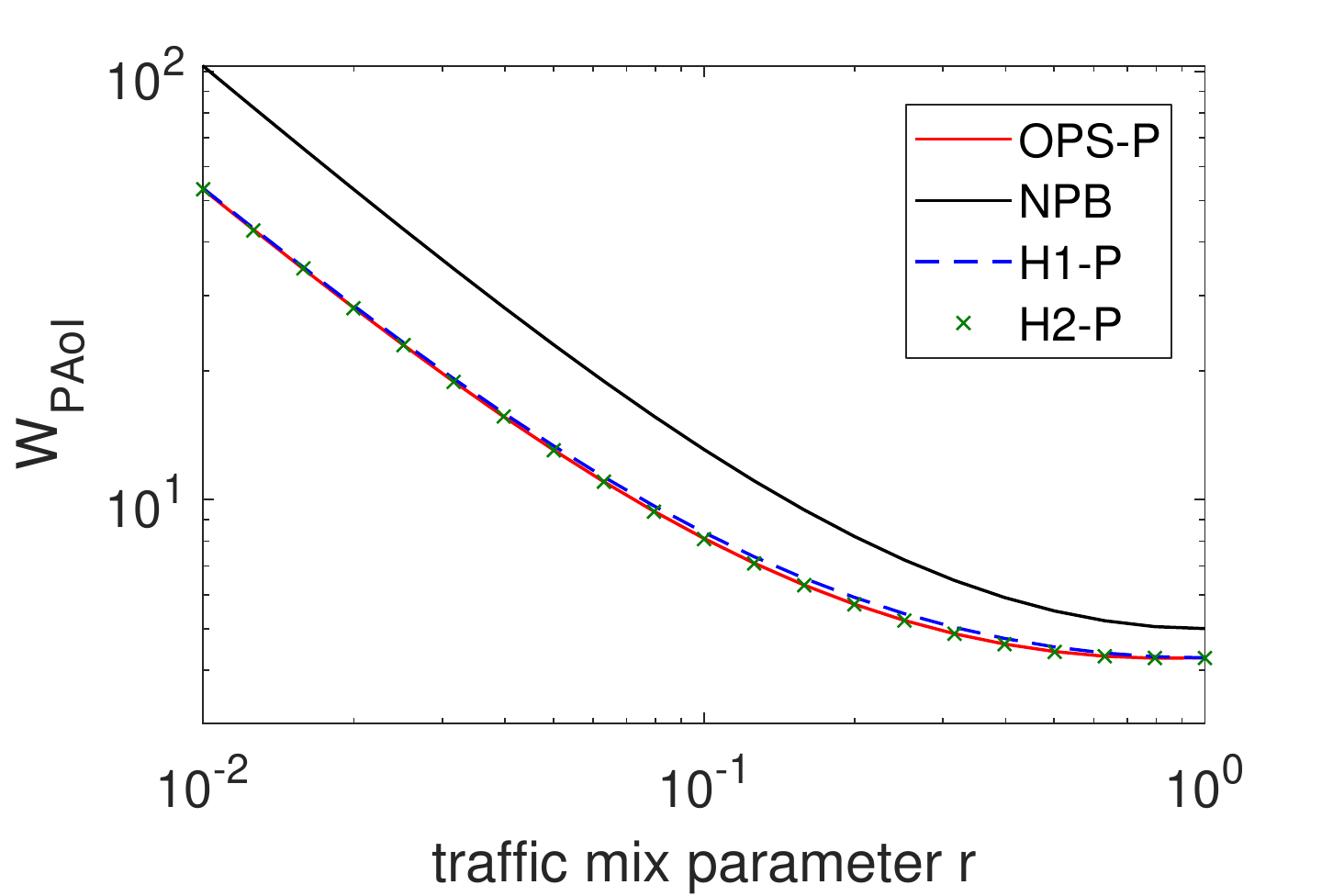}
		\label{fig:symmetric1}
	\end{subfigure}
	\hfill
	\begin{subfigure}[b]{0.48\textwidth}
		\centering
		\caption{$W_{AoI}$ ($\rho=1$)} \vspace*{-0.3cm}
		\includegraphics[width=\textwidth]{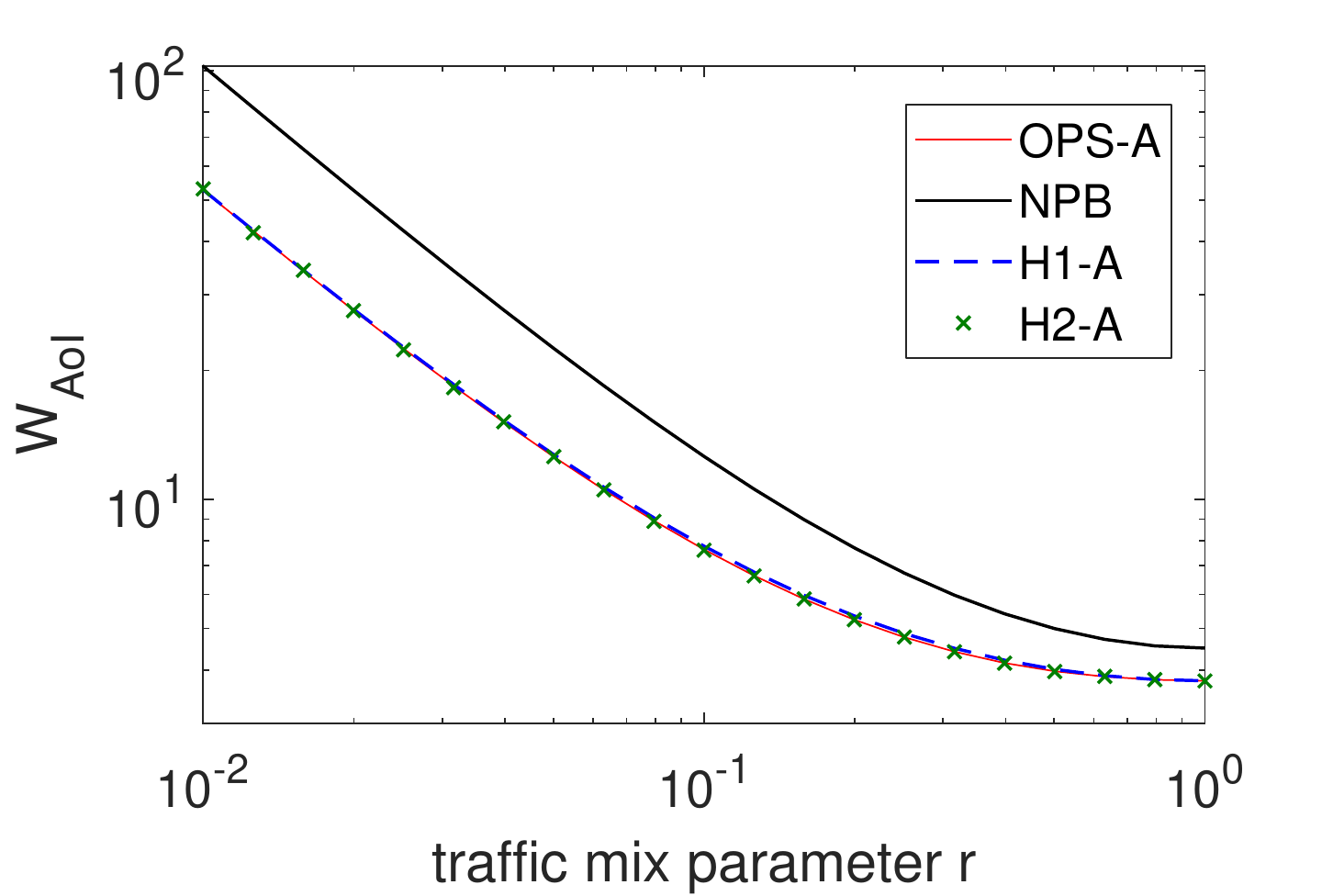}
		\label{fig:symmetric2}
	\end{subfigure}
	\hfill
	\begin{subfigure}[b]{0.48\textwidth}
		\centering
		\caption{$W_{PAoI}$ ($\rho=10$)}  \vspace*{-0.3cm}
		\includegraphics[width=\textwidth]{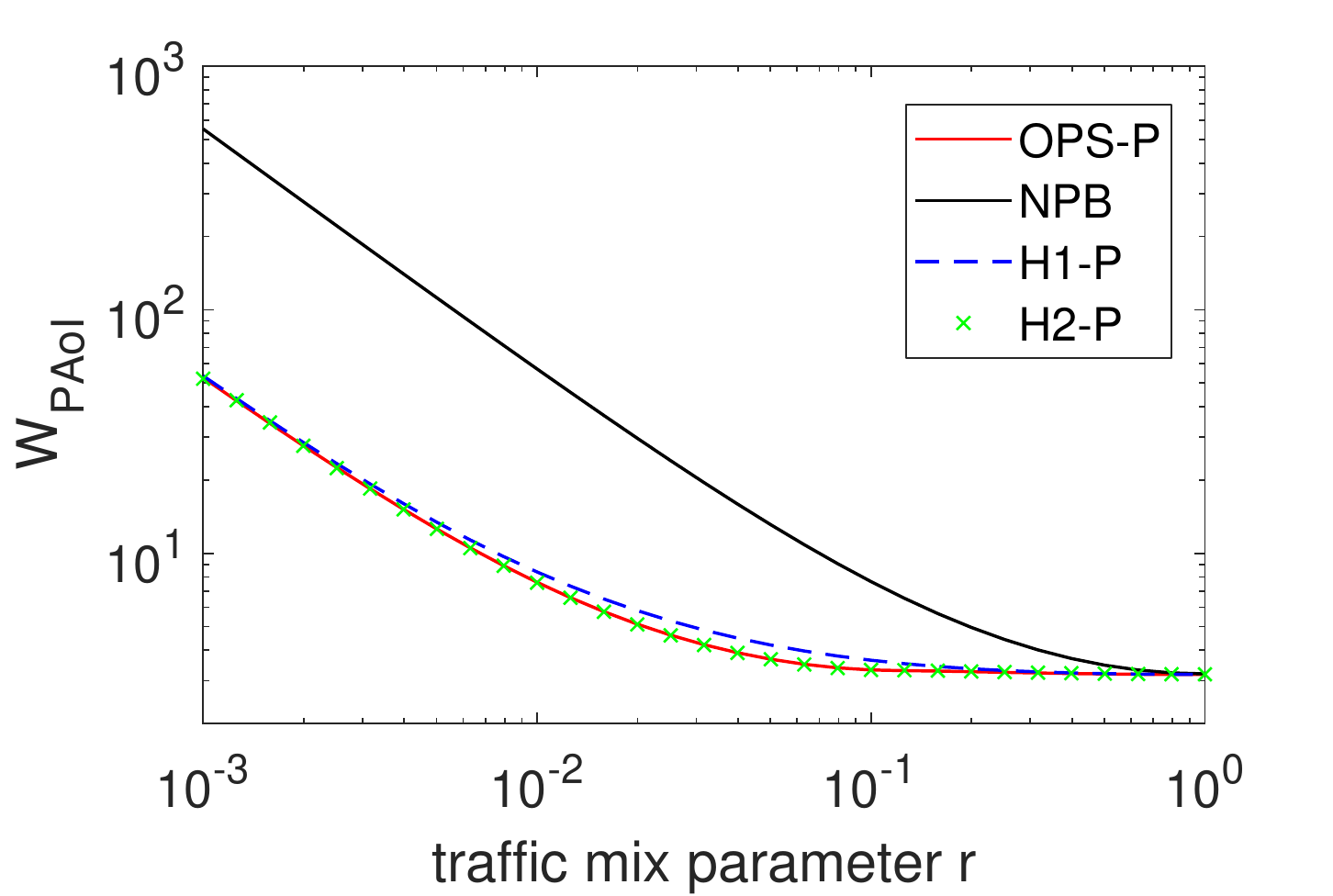}		
		\label{fig:symmetric3}
	\end{subfigure}
	\hfill
	\begin{subfigure}[b]{0.48\textwidth}
		\centering
		\caption{$W_{AoI}$ ($\rho=10$)}  \vspace*{-0.3cm}
		\includegraphics[width=\textwidth]{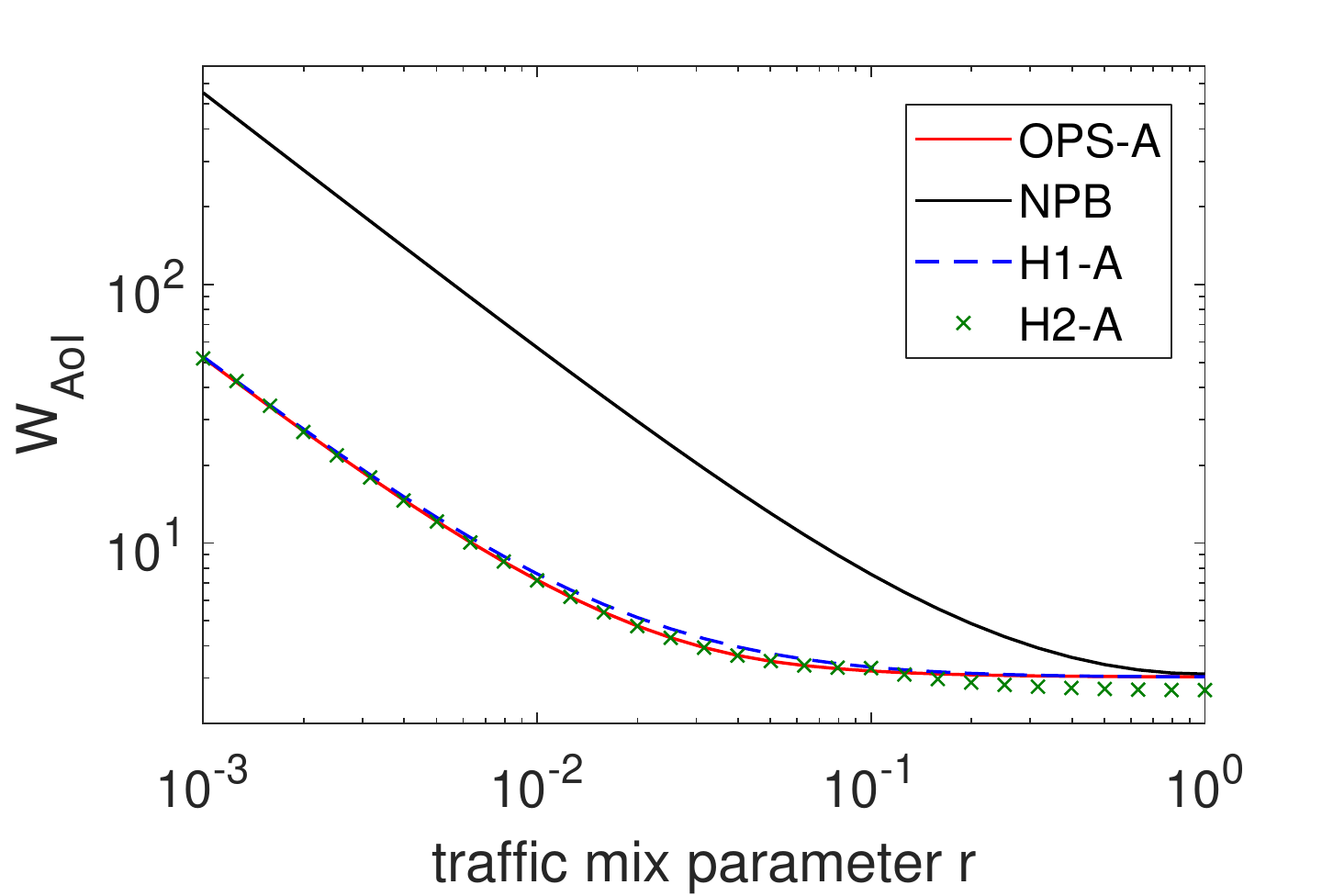}	
		\label{fig:symmetric4}
	\end{subfigure}
	\caption{The weighted average PAoI or AoI as a function the traffic mix parameter $r$ for two values of the load $\rho$.}
	\label{fig:symmetricexample}
\end{figure}

The weighted average PAoI or AoI are depicted in Fig.~\ref{fig:symmetricexample} as a function of the traffix mix parameter $r$ on a log-log scale for two values of the load $\rho$ using the four schedulers OPS-P(A), NPB, H1-P(A), and H2-P(A) and note that H1-P $\equiv$ H1-A and H2-P $\equiv$ H2-A for symmetric networks. We have the following observations about symmetric networks:
\begin{itemize}
	\item For symmetric networks, the optimum traffic mix should be unity due to symmetry. The discrepancy between NPB and the other SBPSQ systems is reduced as $r \rightarrow 1$ and it vanishes as $r \rightarrow 1$ and $\rho \rightarrow \infty$. However, for moderate loads and when $r$ deviates from unity, SBPSQ has substantial advantages compared to NPB.
	\item The proposed heuristic schedulers
	are developed without the knowledge of load and traffic mix
	using only heavy-traffic conditions. However, we observe
	through numerical results that the heuristic schedulers perform very close to that obtained by the computation-intensive optimum probabilistic scheduler. This observation is in line with those made in \cite{joo_eryilmaz_TNET18}.
	\item H2-P (H2-A) presents very similar performance to OPS-P (OPS-A) for all the load and traffic mix values we have obtained whereas H1-P and H1-A are slightly outperformed by them except for light and heavy loads. We also note that there are even cases when H2-A outperforms OPS-A in the high load regime when $r \rightarrow 1$. This stems from the fact that in the heavy-traffic regime, determinized source scheduling strategies perform better than their corresponding probabilistic counterparts for AoI. However, this observation does not necessarily apply to PAoI. 
	\end{itemize}

\subsection{Asymmetric Network - Weighted Average PAoI Minimization}
\begin{figure}
	\centering
	\begin{subfigure}[b]{0.48\textwidth}
		\centering
		\caption{$\omega=4, \ \mu=4, \ r=1$}\vspace*{-0.2cm}
		\includegraphics[width=\textwidth]{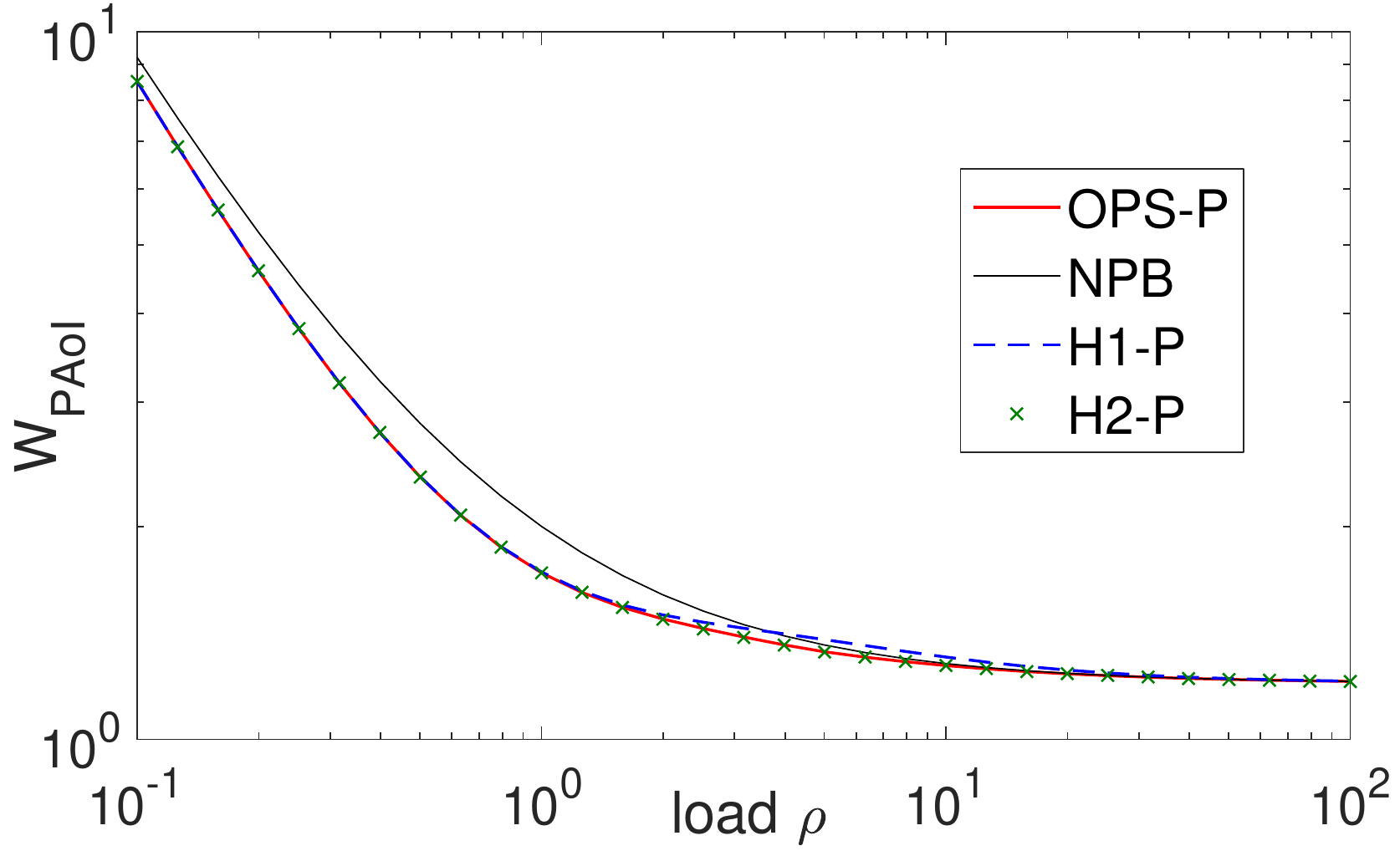}
		%\caption{$y=x$}
		\label{fig:PAoIexample1}
	\end{subfigure}
	\hfill
	\begin{subfigure}[b]{0.48\textwidth}
		\centering
		\caption{$\omega=4, \ \mu=4, \ r=1/4$} \vspace*{-0.2cm}
		\includegraphics[width=\textwidth]{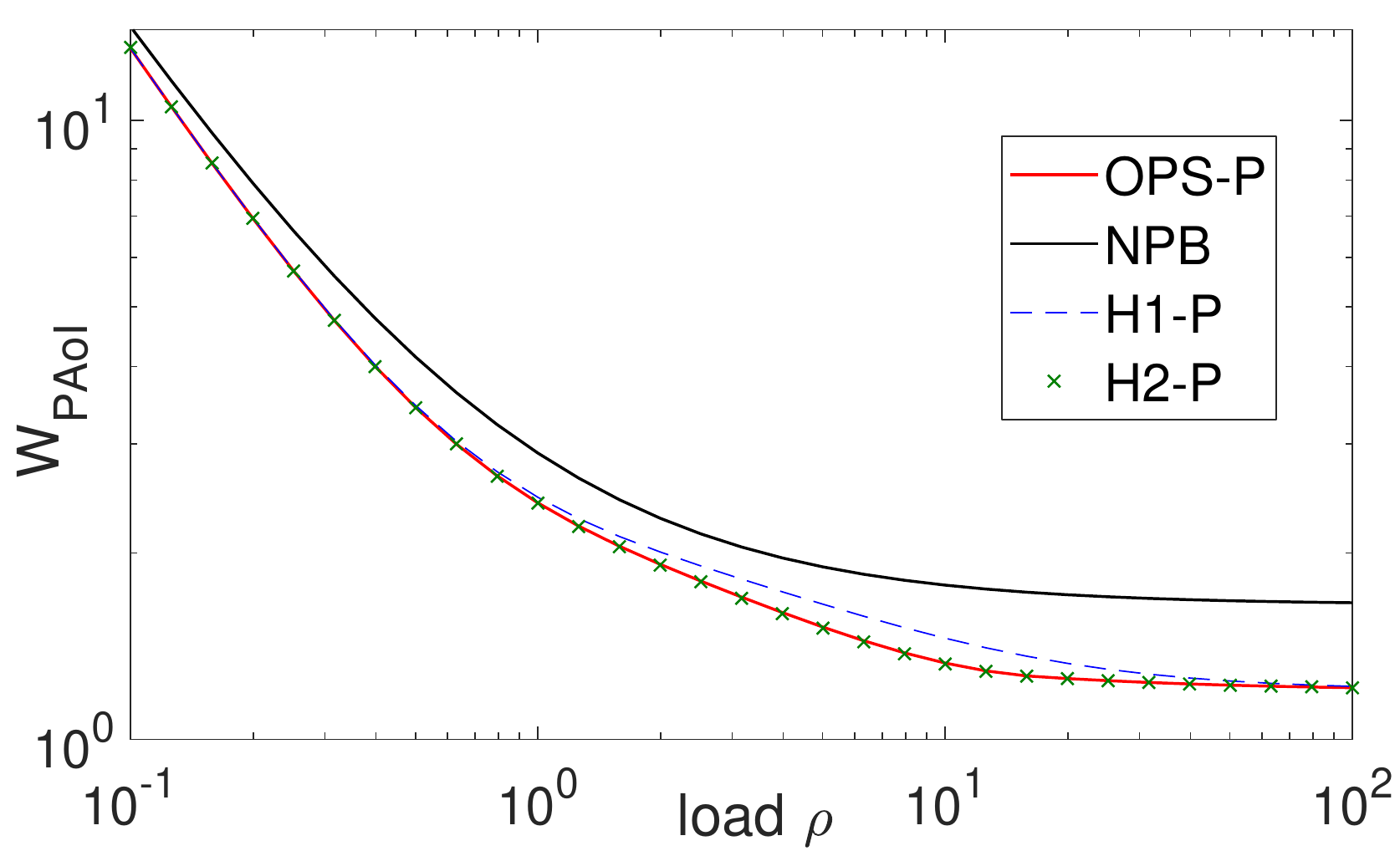}
		%	\caption{$y=3sinx$}
		\label{fig:PAoIexample2}
	\end{subfigure}
	\caption{$W_{PAoI}$ depicted as a function of the total load $\rho$ obtained with the algorithms OPS-P, NPB, H1-P, and H2-P for two different scenarios.}
	%  (a) $\omega=4, \mu=4, \lambda=4, r=1$, (b) $\omega=4, \mu=4, \lambda=1, r=1$.}
	\label{fig:PAoIexample}
\end{figure}

In this numerical example, we depict $W_{PAoI}$ as a function of the load $\rho$ (on a log-log scale) employing four different buffer management/scheduling mechanisms, namely OPS-P, NPB, H1-P, and H2-P in Fig.~\ref{fig:PAoIexample} for which we fix $\omega=4$ and $\mu=4$.   In Fig.~\ref{fig:PAoIexample1}, for given load $\rho$, we choose $\rho_i = \rho r_{PAoI}^*$ where the traffic mix ratio $r_{PAoI}^*=1$ as given in \eqref{optimum_mix_ratio}. This choice ensures that source-$i$ packet generation intensities are chosen such that NBP performance is maximized in terms of $W_{PAoI}$.
On the other hand, for Fig.~\ref{fig:PAoIexample2}, we fix $r=1/4$, a choice which is quite different than the choice $r_{PAoI}^*=1$ giving rise to a scenario for which the arrival rate selections are not as consistent with the weights and average service rates as in Fig.~\ref{fig:PAoIexample1}.
The following observations are made for this example.
\begin{itemize}
	\item If the per-source packet arrival rates are chosen to optimize NBP as in Fig.~\ref{fig:PAoIexample1}, then the discrepancy between NPB and the other three SBPSQ systems is reduced especially for light and heavy loads. For this scenario, there are moderate load values at which NPB outperformed H1-P but OPS-P and H2-P always outperformed NPB in all the cases we had investigated.
	\item When the arrival rates deviate from the the optimum values derived for NPB as in Fig.~\ref{fig:PAoIexample2}, then the advantage of using SBPSQ with respect to NBP is magnified. 
	Therefore, one can conclude that the sensitivity of the performance of SBPSQ systems to the specific choice of the arrival rates are lower than that of NPB.
	\item The performance of H2-P is quite similar to that of OPS-P for all the values we had tried both of which slightly outperform H1-P. We conclude that H2-P depends only on the knowledge on 
	$\omega$ and $\mu$ and does not use the load and traffic mix. However, H2-P can safely be used at all loads and all traffic mixes as a simple-to-implement alternative to OPS-P for weighted PAoI minimization. 
\end{itemize}

 \subsection{Asymmetric Network - Weighted Average AoI Minimization}
\begin{figure}
	\centering
	\begin{subfigure}[b]{0.48\textwidth}
		\centering
		\caption{$\omega=4, \ \mu=4, \ r=1$}\vspace*{-0.2cm}
		\includegraphics[width=\textwidth]{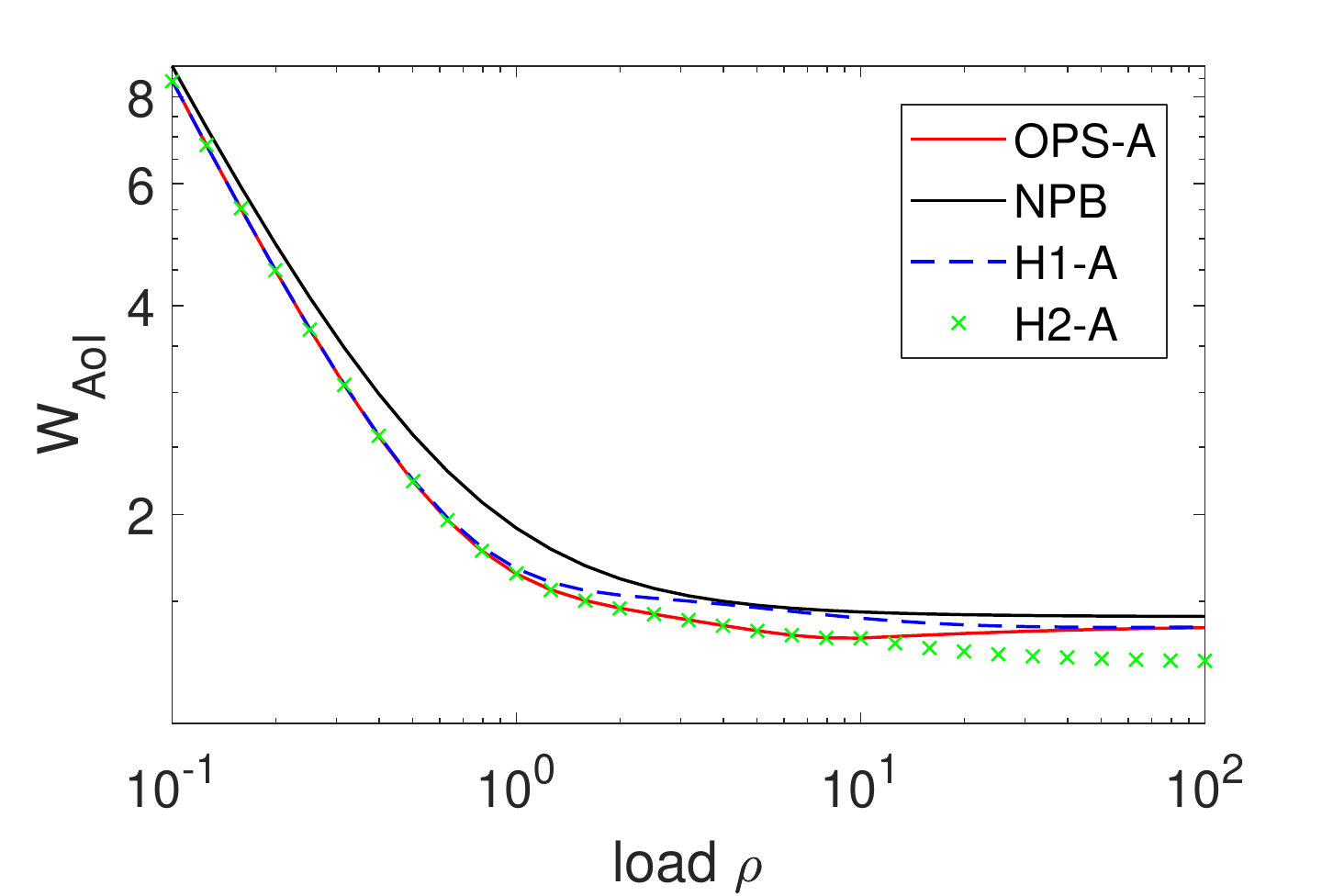}
		%\caption{$y=x$}
		\label{fig:AoIexample1}
	\end{subfigure}
	\hfill
	\begin{subfigure}[b]{0.48\textwidth}
		\centering
		\caption{$\omega=4, \ \mu=4, \ r=1/4$} \vspace*{-0.2cm}
		\includegraphics[width=\textwidth]{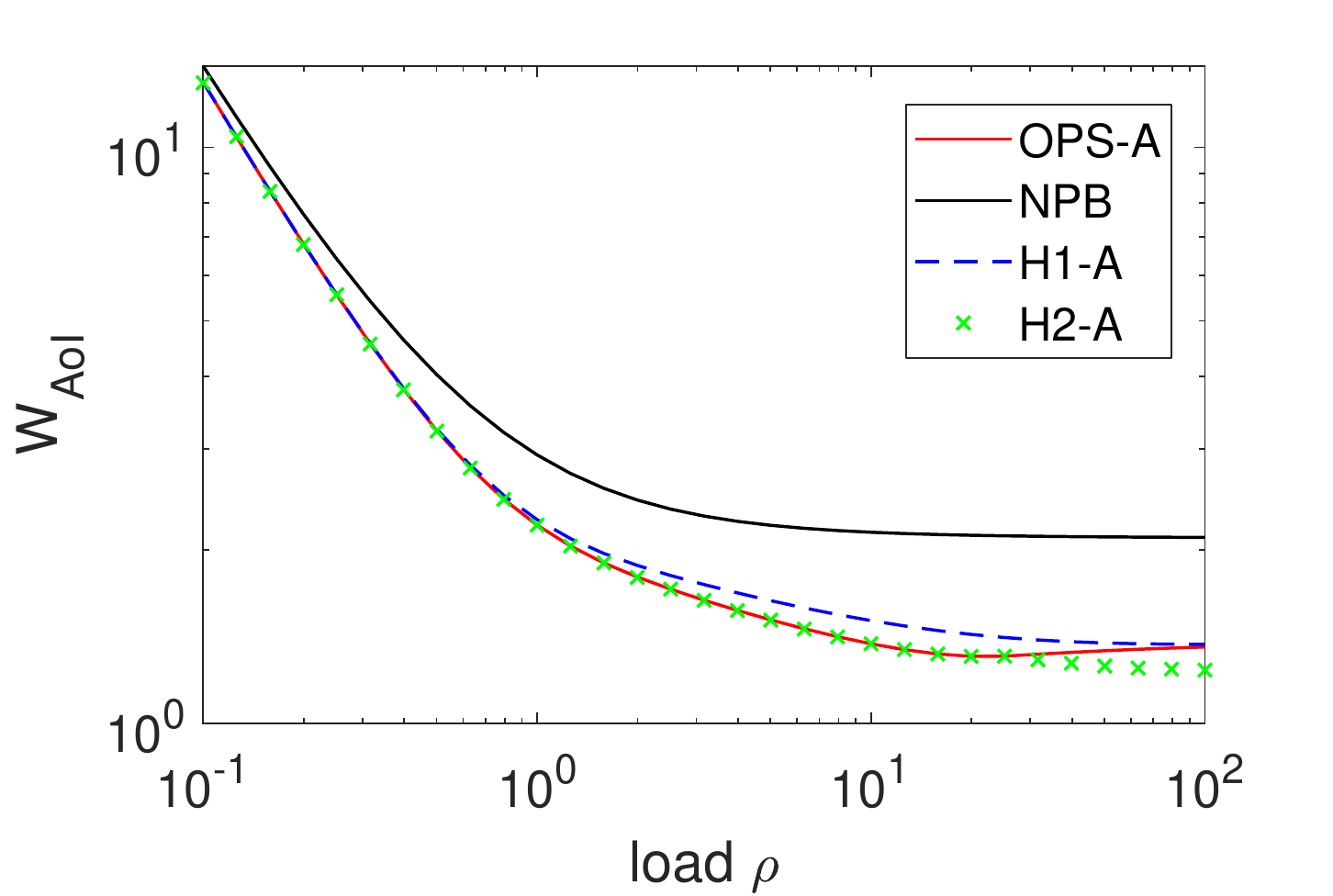}
		%	\caption{$y=3sinx$}
		\label{fig:AoIexample2}
	\end{subfigure}
	\caption{$W_{AoI}$ depicted as a function of the total load $\rho$ obtained with the algorithms OPS-A, NPB, H1-A, and H2-A for two different scenarios.}
	%  (a) $\omega=4, \mu=4, \lambda=4, r=1$, (b) $\omega=4, \mu=4, \lambda=1, r=1$.}
	\label{fig:AoIexample}
\end{figure}

 In this example, we continue with the same example of the previous subsection but we focus on $W_{AoI}$ which is plotted as a function of the load $\rho$ (on a log-log scale) under the policies OPS-A, NPB, H1-A, and H2-A, in Fig.~\ref{fig:AoIexample} with $\omega=4$ and $\mu=4$. The traffic mix 
 parameter $r$ is fixed to $r=1$ and $r=1/4$ in  Fig.~\ref{fig:AoIexample1} and Fig.~\ref{fig:AoIexample2}, respectively.
We have the following observations:
\begin{itemize}
	\item The OPS-A curve is not monotonically decreasing with respect to load $\rho$ as in OPS-P for the two values of the traffic mix parameter $r$ we have studied; it first decreases until a certain load threshold is reached but then it slightly rises up to its heavy-traffic limit obtained with the probability ratio $p_{AoI}^*$. The corresponding load threshold value appears to depend on the traffic mix.
	\item The H2-P policy tracks the performance of OPS-A until the load threshold is reached but when the load ranges between the load threshold and infinity, H2-P outperforms OPS-A. This observation does not pertain to the results obtained for weighted average PAoI minimization. 
\end{itemize}

\section{Conclusions}
\label{section7}
We studied a two-source SBPSQ-based status update system with probabilistic scheduling and we proposed a method to obtain the distributions and moments of AoI and PAoI numerically using CTMCs of absorbing-type. The proposed technique is quite simple to implement making it amenable to use for a wider range of analytical modeling problems regarding AoI/PAoI distributions. Moreover, we performed heavy-traffic analysis for the same scenario to obtain closed form expressions for the per-source average AoI/PAoI values from which we have proposed two simple-to-implement age-agnostic heuristic schedulers.
The proposed heuristic schedulers
are developed without the knowledge of load and traffic mix
using only heavy-traffic conditions. However, we observed
through numerical results that the heuristic schedulers perform very close to that obtained by their computation-intensive optimum probabilistic scheduler counterparts and at all loads and traffic mixes. In particular, for weighted AoI minimization, our proposed heuristic scheduler H2-A's performance tracked that of the optimum probabilistic scheduler OPS-A except for heavy loads where it even outperformed OPS-A.  For weighted PAoI minimization, our proposed heuristic scheduler H2-P's performance tracked that of the optimum probabilistic scheduler OPS-P. Therefore, H2-A and H2-P are promising candidates for scheduling in SBPSQ systems stemming from their performance and age-agnostic nature. Future work will be on extending the results to general number of sources and  non-exponentially distributed service times, and also to discrete-time.

% or


%\appendices
%% you can choose not to have a title for an appendix
%% if you want by leaving the argument blank
%\section{}

%\section*{Acknowledgment}

\bibliographystyle{unsrtnat}

\end{document}